\documentclass[aps,prl,twocolumn,showpacs,superscriptaddress,nofootinbib]{revtex4-2}

\usepackage{amsmath,amssymb}
\usepackage{graphicx}
\usepackage{hyperref}
\usepackage{physics}
\usepackage{siunitx}
\usepackage{enumitem}

\begin{document}

\title{First Observation of Dispersive Shock Waves in an Electron Beam}

\author{H. McCright}
\affiliation{Department of Physics, University of Maryland, College Park, US}

\author{I.G. Abel}
\affiliation{Institute for Research in Electronics and Applied Physics, University of Maryland, College Park, US}

\author{I. Haber}
\affiliation{Institute for Research in Electronics and Applied Physics, University of Maryland, College Park, US}

\author{P.G. O'Shea}
\affiliation{Department of Electrical and Computer Engineering, University of Maryland, College Park, US}

\author{B.L. Beaudoin}
\affiliation{Institute for Research in Electronics and Applied Physics, University of Maryland, College Park, US}

\date{\today}

\begin{abstract}
Dispersive shock waves (DSWs) are expanding nonlinear wave trains that arise when dispersion regularizes a steepening front, a phenomenon observed in fluids, plasmas, optics, and superfluids. Here we report the first experimental observation of DSWs in an intense electron beam, using the University of Maryland Electron Ring (UMER). A localized induction-cell perturbation produced a negative density pulse that evolved into a leading soliton-like peak followed by an expanding train of oscillations. The leading peak satisfied soliton scaling laws for $(width)^{2}$ vs inverse amplitude and velocity vs amplitude, while the total wave-train width increased linearly with time, consistent with Korteweg--de Vries (KdV) predictions. Successive peaks showed decreasing amplitude and velocity toward the trailing edge, in agreement with dispersive shock ordering. These results demonstrate that intense charged particle beams provide a new laboratory platform for studying dispersive hydrodynamics, extending nonlinear wave physics into the high-intensity beam regime.
\end{abstract}

\maketitle

\noindent


Dispersive shock waves (DSWs) are expanding nonlinear wave trains that regularize a steep front through the balance of dispersion and nonlinearity \cite{PhysRevLett.116.174501}. They appear widely in nature, from tidal bores to atmospheric “morning glory” waves \cite{PhysRevX.4.021022}, and have been reproduced in laboratory systems including plasmas \cite{Taylor1970, Niemann2014, 10.1063/1.1693287, Craig_1974}, nonlinear optics \cite{Wan2007, Jia2007, Ghofraniha2007, PhysRevX.4.021022}, classical fluids \cite{Smyth1988, Lighthill1978}, and superfluids \cite{Dutton2001, PhysRevLett_Chang}. Across these different media, their defining signature is consistent: a sharp leading edge followed by an oscillatory wake.

Charged particle beams provide a new platform for nonlinear dispersive dynamics. In particular, space-charge–dominated (“intense”) beams, where collective forces outweigh emittance in the transverse envelope equation \cite{reiser1994}, support collective behaviors analogous to fluids \cite{4331740}. Previous work at the University of Maryland Electron Ring (UMER) demonstrated Korteweg–de Vries (KdV) solitons, which are localized, self-reinforcing waves that maintain their shape through a balance of nonlinearity and dispersion, launched by localized peaks in beam current \cite{charles,Mo}.

Here, we report the first observation of DSWs in an intense electron beam. Unlike solitons, which remain localized, DSWs form from a steep disturbance and expand into a train of oscillations that collectively smooth the front. In UMER, these DSWs are triggered by a dip in beam current and evolve into a high-frequency train of oscillations, representing a qualitatively distinct nonlinear mechanism. Long  propagation distances and highly reproducible beam conditions enable controlled studies of DSW formation with long-time evolution. By generating both peak-driven solitons and dip-driven DSWs in a reproducible setting, these experiments provide a platform for investigating nonlinear wave interactions in beams, as recent theory has highlighted the rich dynamics of DSW collisions \cite{cdvf-xnfw}. Because beams in high-energy accelerators also start in the space-charge–dominated regime, the dynamics reported here are not confined to low-energy models: they are a general feature of intense beam physics, with implications for accelerator design and astrophysical plasmas \cite{ElTantawy2016, Shan2024}.

The dynamics of small-amplitude density perturbations in a dispersive, nonlinear medium are captured by the KdV equation \cite{Korteweg01051895}:

\begin{equation}
\frac{\partial u}{\partial t} + \alpha u \frac{\partial u}{\partial z} + \beta \frac{\partial^3 u}{\partial z^3} = 0 ,
\end{equation}
where $u(z,t)$ is the density perturbation, $\alpha$ is the nonlinear coefficient, and $\beta$ governs dispersion. Historically, the KdV equation admits solitary and cnoidal wave solutions. Gurevich and Pitaevskii \cite{Gurevich1974} solved the nonstationary KdV Riemann problem, showing that a step-like initial perturbation evolves into a modulated wave train. Numerical studies \cite{Fornberg1978} confirmed that the fastest, largest-amplitude peak leads the train, a key feature of a DSW.

\begin{figure}[h]
    \centering
    \includegraphics[width=0.9\columnwidth]{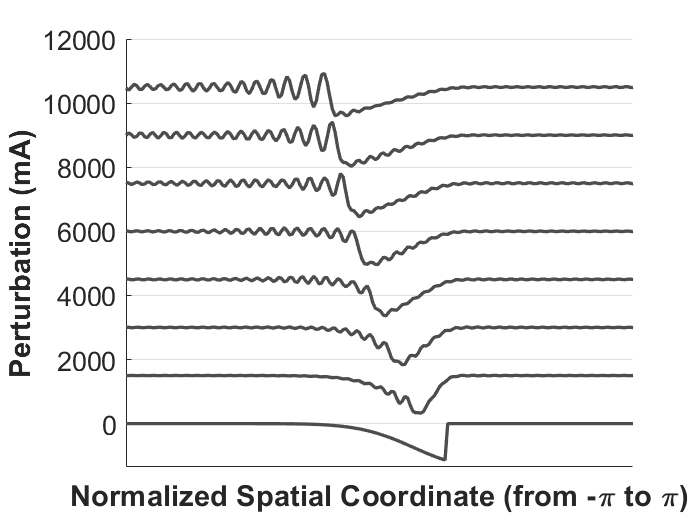}
    \caption{Numerical KdV solution for a negative initial perturbation on a periodic domain $x \in [-\pi, \pi]$, computed using a pseudo-spectral spatial method and a fourth-order Runge--Kutta time stepper. Profiles are shown in $100\ \mu\text{s}$ increments, increasing upward in time.}
    \label{kdv}
\end{figure}

For positive pulses, the nonlinear steepening is balanced by dispersion, producing solitons; this is the regime of previously observed solitons in UMER \cite{charles,Mo}. For negative perturbations, however, the steepening leads to a gradient catastrophe followed by dispersion-dominated regularization, resulting in an expanding oscillatory front rather than a solitary peak (Fig.~\ref{kdv}). 

Physically, faster characteristics from larger-amplitude regions overtake slower ones, steepening the front until dispersion regularizes it into an oscillatory train \cite{Gurevich1974, Fornberg1978}. A pseudopotential analysis \cite{lyu2005} confirms that negative perturbations do not support localized solitons but allow oscillatory trajectories, producing peaks whose amplitude and velocity decrease toward the trailing edge.

To identify a DSW experimentally, we therefore test for the following:
\begin{enumerate}[itemsep=0.5em, topsep=0.5em, leftmargin=*]
    \item Soliton-like leading peak: The first peak should follow the linear scaling laws for solitons measured previously in UMER (velocity vs amplitude) and $(width)^{2}$ vs $1/\text{amplitude}$. \cite{Mo}
    
    \item Two distinct edge velocities: The leading and trailing edges of the DSW propagate with different speeds $s_+$ and $s_-$, predicted for KdV DSWs with steep initial conditions as
    \begin{equation}
    s_- = -\Delta + u_+, \quad s_+ = \frac{2}{3}\Delta + u_+,
    \end{equation}
    where $\Delta = u_+ - u_-$ is the initial amplitude difference across the shock. The DSW therefore expands at a constant rate
    \begin{equation}
    W(t) = (s_+ - s_-)t = \frac{5}{3}\Delta\,t,
    \end{equation}
    implying a linear growth of the DSW width with slope proportional to the initial amplitude difference across the shock \cite{Gurevich1974,el2016dispersive}.
    
\end{enumerate}

We studied DSW formation using the 10-keV electron beam in the 11.52m circumference UMER storage ring, both experimentally and via simulations in the WARP particle-in-cell code \cite{GROTE1996193}. In WARP, the beam was initialized with uniform density and transverse velocity, with a spread much smaller than the beam speed ($\Delta v \ll \beta c$, $\beta \approx v_\mathrm{beam}/c$), using 16 million macroparticles—computational particles each representing many electrons—along with a 1~ns time step, 64 radial cells, and 2048 axial cells. The domain spanned 0.0254~m radially (conducting) and 11.52~m axially (periodic), with parameters verified for convergence and prior experimental validation \cite{Mo}.
 
\begin{figure}[h]
\centering
\includegraphics[width=\columnwidth]{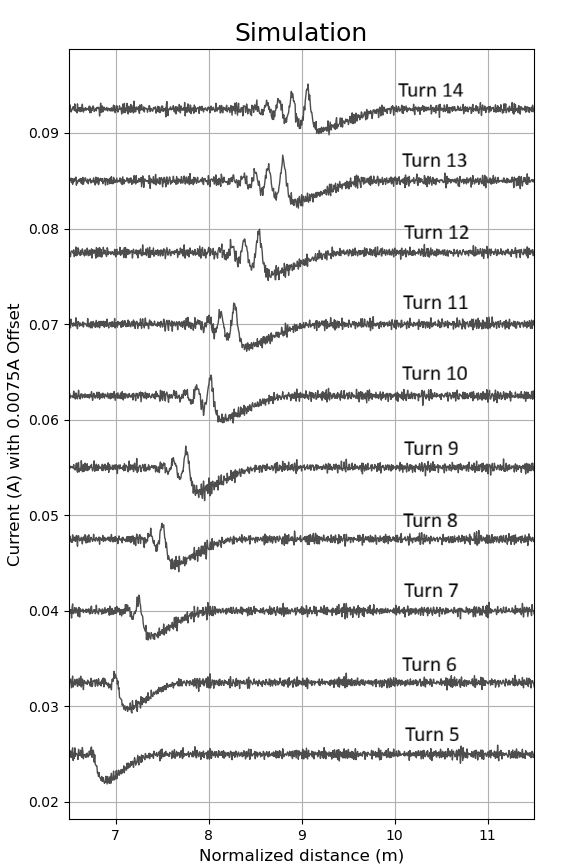}
\caption{Simulated beam current evolution using WARP for turns 5--14, stacked with 0.0075~A upward offsets. Early turns are omitted while the beam relaxes in phase space; the DSW forms once the initial negative density perturbation steepens.}
\label{warp}
\end{figure}

In WARP, a localized, one-time longitudinal electric field introduced a narrow Gaussian velocity perturbation (10ns travel time). After a single pass, the perturbation was removed, producing two opposite-polarity current peaks that split into slow and fast waves \cite{PhysRevLett.71.1836, PhysRevSTAB.13.034201}, eventually forming dispersive shock structures (\autoref{warp}). The simulation illustrates the qualitative evolution of a leading wave followed by smaller oscillations.

Experimentally, a single 100ns rectangular bunch was injected, with current measured per turn using a wall current monitor 7.67m downstream. A controlled velocity modulation was applied via an inline induction cell, producing a negative density perturbation consistent with the simulation \cite{beaudoin2008}.

\begin{figure}[h]
    \centering
    \includegraphics[width=\columnwidth]{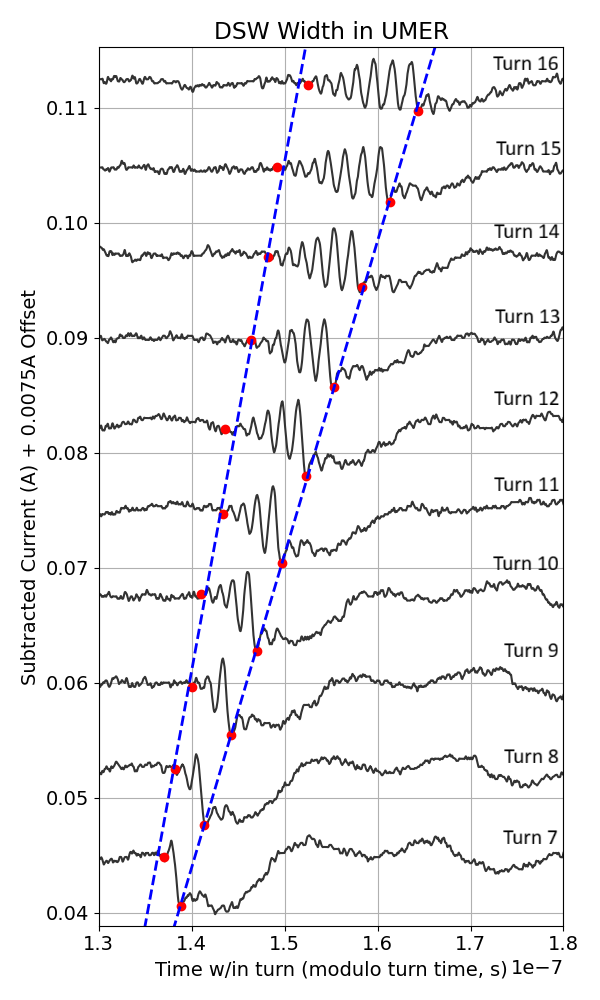}
    \caption{Measured current perturbation (A, background-subtracted) vs time (s) at successive revolutions in UMER, showing the formation of a dispersive shock wave from an initial negative density perturbation. The initial beam current was approximately 30~mA. Turn number increases from bottom to top. The width of the DSW at each turn was measured between the red dots, with the blue dotted lines fitted to these points to illustrate the linear expansion of the DSW.}
    \label{data}
\end{figure}

\autoref{data} shows the measured evolution of the current perturbation over successive turns 5--16. The data are background-subtracted using the beam when perturbations are not intentionally introduced to highlight the oscillatory behavior. No Fourier cutoff or other frequency filtering was applied, in order to preserve the full range of oscillations and avoid inadvertently removing or distorting features of the DSW. A fully developed DSW pattern emerges after $\sim 10$ revolutions, with a high-amplitude, fast-moving leading peak followed by a trailing oscillatory train.

We restricted the quantitative analysis to turns 10--16. At earlier turns, the perturbation is still evolving toward a steady DSW structure; the leading peak and subsequent oscillations are not yet well separated. At later turns, beam loss and phase-space dilution reduce signal-to-noise, systematically biasing amplitude and velocity measurements, particularly for the trailing peaks.

\begin{figure}[h]
    \centering
    \includegraphics[width=\columnwidth]{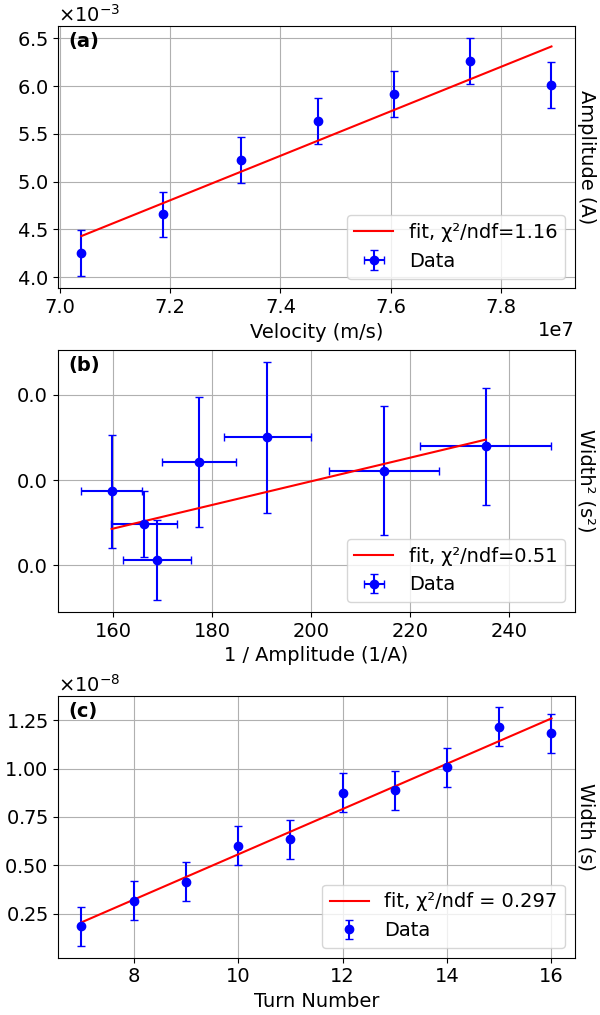}
    \caption{Three-panel figure illustrating key features of the dispersive shock wave data:\\
    a) Squared width of the leading peak vs inverse amplitude, showing linear scaling consistent with soliton behavior. Reduced $\chi^2$ = 0.51.\\
    b) Velocity of the leading peak vs amplitude, showing linear scaling consistent with soliton behavior. Reduced $\chi^2$ = 1.16.\\
    c) DSW width versus turn number, showing linear expansion. Width as measured between blue lines in \autoref{data}. Width of DSW expands linearly, as predicted with a reduced $\chi^2$ of 0.297.
    }
    \label{PLOT}
\end{figure}

The leading peak obeyed the soliton scaling laws established in prior measurements:
(i) velocity vs amplitude (\autoref{PLOT}a) and  
(ii) $(width)^{2}$ vs $1/\text{amplitude}$ (\autoref{PLOT}b). The soliton width was defined as the full width at half maximum (FWHM) of the leading density peak.
In both cases, a linear fit produced reduced $\chi^2$ values of 0.51 and 1.16, respectively, suggesting that the data is well described by the linear models.

The separation between the leading and trailing edges increased linearly with turn number, as shown in \autoref{PLOT}c, consistent with the KdV prediction $W(t)~\propto~(u_-~- u_+)t$. Both linear fits yield reduced chi-squared values near unity, indicating excellent agreement with a linear model. For the data shown in \autoref{data}, the inferred expansion rate of the DSW width was 
$\mathrm{d}W/\mathrm{dt} = (6.02 \pm 0.31)\times 10^{-3}~\mathrm{s/s}$.

To probe the amplitude dependence of $W$, separate experiments were carried out with a smaller initial perturbation, implemented by reducing the magnitude of the applied velocity modulation from 30~mA to 20~mA. In these experiments, the inferred spreading rate was 
$\mathrm{d}W/\mathrm{dt} = (4.04 \pm 0.33) \times 10^{-3}~\mathrm{s/s}$, 
and the initial amplitude ratio was roughly 0.75. This is consistent with the linear scaling predicted~\cite{Gurevich1974,el2016dispersive}.

Successive peaks exhibit decreasing amplitude and velocity toward the trailing edge, consistent with the expected ordering in a DSW. This is reflected in \autoref{PLOT}c by the smaller slope in the DSW width between the first and second peaks compared to the slope between the first and third peaks.

We report the first observation of DSWs in a space-charge–dominated electron beam. Controlled negative density perturbations in UMER generated expanding nonlinear wave trains whose leading peak follows soliton-like scaling laws, while the total wave-train width grows linearly with time, in agreement with KdV theory. These results extend prior soliton studies into a distinct nonlinear regime, demonstrating that DSW dynamics are accessible in intense charged particle beams.

UMER’s storage-ring geometry and reproducible beam conditions enable detailed tracking of DSW formation. Beyond fundamental interest, these findings open the door to studies of soliton–DSW interactions, modulational instabilities, and combined excitations. Because high-energy accelerators also begin in a space-charge–dominated regime, the observed dynamics are scalable, establishing intense electron beams as a versatile platform for nonlinear dispersive physics with potential relevance to accelerator science, beam-driven radiation sources, and astrophysical plasma analogues.

\begin{acknowledgments}
Thanks to Carter Hall for the opportunity to pursue this project as part of H. McCright's Physics Undergraduate Honors Thesis at the University of Maryland, College Park, and Joseph Zennamo at Fermilab for guidance, mentorship, and feedback on drafts. Special thanks to Rollo the cat, a steadfast source of support throughout the project. This work was supported by DOE Grant No.DE-SC0022009.
\end{acknowledgments}

\bibliographystyle{apsrev4-2}
\bibliography{refs.bib}

\end{document}